\documentclass[12pt,a4paper]{article}

\usepackage{ifthen} 

\newboolean{articletitles}
\setboolean{articletitles}{true} 

\newboolean{uprightparticles}
\setboolean{uprightparticles}{false} 

\usepackage{rotating}  

\usepackage{lineno}  
\usepackage{xspace} 
\usepackage{graphicx}  
\usepackage{color}
\usepackage{colortbl}
\graphicspath{{./figs/}} 
\usepackage{amsmath} 
\usepackage{amssymb}
\usepackage{amsfonts}
\usepackage{upgreek} 
\newcommand*\patchAmsMathEnvironmentForLineno[1]{%
\expandafter\let\csname old#1\expandafter\endcsname\csname #1\endcsname
\expandafter\let\csname oldend#1\expandafter\endcsname\csname
end#1\endcsname
\renewenvironment{#1}%
  {\linenomath\csname old#1\endcsname}%
  {\csname oldend#1\endcsname\endlinenomath}%
}
\newcommand*\patchBothAmsMathEnvironmentsForLineno[1]{%
  \patchAmsMathEnvironmentForLineno{#1}%
  \patchAmsMathEnvironmentForLineno{#1*}%
	   }
\AtBeginDocument{%
\patchBothAmsMathEnvironmentsForLineno{equation}%
\patchBothAmsMathEnvironmentsForLineno{align}%
\patchBothAmsMathEnvironmentsForLineno{flalign}%
\patchBothAmsMathEnvironmentsForLineno{alignat}%
\patchBothAmsMathEnvironmentsForLineno{gather}%
\patchBothAmsMathEnvironmentsForLineno{multline}%
}

\usepackage{hyperref}    
\usepackage[all]{hypcap} 
\def\lhcb {\mbox{LHCb}\xspace}

\ifthenelse{\boolean{uprightparticles}}%
{

 \def\Ppi         {\ensuremath{\uppi}\xspace}

 \def\PDelta      {\ensuremath{\Delta}\xspace}                 
 \def\PXi      {\ensuremath{\Xi}\xspace}                 
 \def\PLambda      {\ensuremath{\Lambda}\xspace}                 
 \def\PSigma      {\ensuremath{\Sigma}\xspace}                 
 \def\POmega      {\ensuremath{\Omega}\xspace}                 
 \def\PUpsilon      {\ensuremath{\Upsilon}\xspace}

 \def\PB      {\ensuremath{\mathrm{B}}\xspace}                 
                  
 \def\PD      {\ensuremath{\mathrm{D}}\xspace}

 \def\PK      {\ensuremath{\mathrm{K}}\xspace}

 \def\Pb      {\ensuremath{\mathrm{b}}\xspace}                 
 \def\Pc      {\ensuremath{\mathrm{c}}\xspace}

 \def\Pi      {\ensuremath{\mathrm{i}}\xspace}

 \def\Pp      {\ensuremath{\mathrm{p}}\xspace}

 \def\Ps      {\ensuremath{\mathrm{s}}\xspace}

}
{

 \def\Ppi         {\ensuremath{\pi}\xspace}

 \mathchardef\PDelta="7101
 \mathchardef\PXi="7104
 \mathchardef\PLambda="7103
 \mathchardef\PSigma="7106
 \mathchardef\POmega="710A
 \mathchardef\PUpsilon="7107
                  
 \def\PB      {\ensuremath{B}\xspace}                 
                  
 \def\PD      {\ensuremath{D}\xspace}

 \def\PK      {\ensuremath{K}\xspace}

 \def\Pb      {\ensuremath{b}\xspace}                 
 \def\Pc      {\ensuremath{c}\xspace}

 \def\Pi      {\ensuremath{i}\xspace}

 \def\Pp      {\ensuremath{p}\xspace}

 \def\Ps      {\ensuremath{s}\xspace}

}

\def\squark    {\ensuremath{\Ps}\xspace}

\def\cquark    {\ensuremath{\Pc}\xspace}

\def\bquark    {\ensuremath{\Pb}\xspace}

\def\pion  {\ensuremath{\Ppi}\xspace}

\def\pip   {\ensuremath{\pion^+}\xspace}
\def\pim   {\ensuremath{\pion^-}\xspace}
\def\pipi  {\ensuremath{\pion^+\pion^-}\xspace}
\def\pipm  {\ensuremath{\pion^\pm}\xspace}

\def\kaon  {\ensuremath{\PK}\xspace}
  \def\Kbar  {\kern 0.2em\overline{\kern -0.2em \PK}{}\xspace}

\def\Kz    {\ensuremath{\kaon^0}\xspace}
\def\Kzb   {\ensuremath{\Kbar^0}\xspace}
\def\KzKzb {\ensuremath{\Kz \kern -0.16em \Kzb}\xspace}
\def\Kp    {\ensuremath{\kaon^+}\xspace}
\def\Km    {\ensuremath{\kaon^-}\xspace}
\def\Kpm   {\ensuremath{\kaon^\pm}\xspace}

\def\KpKm  {\ensuremath{\Kp \kern -0.16em \Km}\xspace}

  \def\Dbar    {\kern 0.2em\overline{\kern -0.2em \PD}{}\xspace}
\def\D       {\ensuremath{\PD}\xspace}

\def\Dz      {\ensuremath{\D^0}\xspace}
\def\Dzb     {\ensuremath{\Dbar^0}\xspace}
\def\DzDzb   {\ensuremath{\Dz {\kern -0.16em \Dzb}}\xspace}
\def\Dp      {\ensuremath{\D^+}\xspace}
\def\Dm      {\ensuremath{\D^-}\xspace}

\def\DpDm    {\ensuremath{\Dp {\kern -0.16em \Dm}}\xspace}
\def\Dstar   {\ensuremath{\D^*}\xspace}

\def\Dstarp  {\ensuremath{\D^{*+}}\xspace}
\def\Dstarm  {\ensuremath{\D^{*-}}\xspace}

\def\Dstarmp {\ensuremath{\D^{*\mp}}\xspace}
\def\Ds      {\ensuremath{\D^+_\squark}\xspace}
\def\Dsp     {\ensuremath{\D^+_\squark}\xspace}
\def\Dsm     {\ensuremath{\D^-_\squark}\xspace}

\def\B       {\ensuremath{\PB}\xspace}
  \def\Bbar    {\kern 0.18em\overline{\kern -0.18em \PB}{}\xspace}

\def\Bu      {\ensuremath{\B^+}\xspace}

\def\Bp      {\ensuremath{\Bu}\xspace}

\def\Bpm     {\ensuremath{\B^\pm}\xspace}

\def\Bd      {\ensuremath{\B^0}\xspace}
\def\Bs      {\ensuremath{\B^0_\squark}\xspace}
\def\Bds     {\ensuremath{\B^0_{(\squark)}}\xspace}
\def\Bsb     {\ensuremath{\Bbar^0_\squark}\xspace}

  \def\Y#1S{\ensuremath{\PUpsilon{(#1S)}}\xspace}

\def\proton      {\ensuremath{\Pp}\xspace}

\def\L {\ensuremath{\PLambda}\xspace}
\def\Lbar {\ensuremath{\kern 0.1em\overline{\kern -0.1em\PLambda}}\xspace}

\def\Lb      {\ensuremath{\L^0_\bquark}\xspace}

\def\to                 {\ensuremath{\rightarrow}\xspace}

\def\CP                {\ensuremath{C\!P}\xspace}

\def\AT#1     {\ensuremath{A_{\mathrm{T}}^{#1}}\xspace}           

\def\C#1      {\ensuremath{\mathcal{C}_{#1}}\xspace}                       
\def\Cp#1     {\ensuremath{\mathcal{C}_{#1}^{'}}\xspace}                    
\def\Ceff#1   {\ensuremath{\mathcal{C}_{#1}^{\mathrm{(eff)}}}\xspace}        
\def\Cpeff#1  {\ensuremath{\mathcal{C}_{#1}^{'\mathrm{(eff)}}}\xspace}       
\def\Ope#1    {\ensuremath{\mathcal{O}_{#1}}\xspace}                       
\def\Opep#1   {\ensuremath{\mathcal{O}_{#1}^{'}}\xspace}                    

\newcommand{\tev}{\ensuremath{\mathrm{\,Te\kern -0.1em V}}\xspace}
\newcommand{\gev}{\ensuremath{\mathrm{\,Ge\kern -0.1em V}}\xspace}
\newcommand{\mev}{\ensuremath{\mathrm{\,Me\kern -0.1em V}}\xspace}
\newcommand{\kev}{\ensuremath{\mathrm{\,ke\kern -0.1em V}}\xspace}
\newcommand{\ev}{\ensuremath{\mathrm{\,e\kern -0.1em V}}\xspace}
\newcommand{\gevc}{\ensuremath{{\mathrm{\,Ge\kern -0.1em V\!/}c}}\xspace}
\newcommand{\mevc}{\ensuremath{{\mathrm{\,Me\kern -0.1em V\!/}c}}\xspace}
\newcommand{\gevcc}{\ensuremath{{\mathrm{\,Ge\kern -0.1em V\!/}c^2}}\xspace}
\newcommand{\gevgevcccc}{\ensuremath{{\mathrm{\,Ge\kern -0.1em V^2\!/}c^4}}\xspace}
\newcommand{\mevcc}{\ensuremath{{\mathrm{\,Me\kern -0.1em V\!/}c^2}}\xspace}

\def\mum  {\ensuremath{\,\upmu\rm m}\xspace}

\def\invfb   {\ensuremath{\mbox{\,fb}^{-1}}\xspace}

\newcommand{\chisq}{\ensuremath{\chi^2}\xspace}

\def\gsim{{~\raise.15em\hbox{$>$}\kern-.85em
          \lower.35em\hbox{$\sim$}~}\xspace}
\def\lsim{{~\raise.15em\hbox{$<$}\kern-.85em
          \lower.35em\hbox{$\sim$}~}\xspace}

\def\pt         {\mbox{$p_{\rm T}$}\xspace}

\def\rad{\ensuremath{\rm \,rad}\xspace}

\def\evtgen     {\mbox{\textsc{EvtGen}}\xspace}
\def\pythia     {\mbox{\textsc{Pythia}}\xspace}

\def\geant      {\mbox{\textsc{Geant4}}\xspace}

\def\gauss      {\mbox{\textsc{Gauss}}\xspace}

\def\tell1  {TELL1\xspace}
\def\ukl1   {UKL1\xspace}

\usepackage{cite} 
\usepackage{mciteplus}

\begin{document}

\renewcommand{\thefootnote}{\fnsymbol{footnote}}
\setcounter{footnote}{1}

\begin{titlepage}
\pagenumbering{roman}

\vspace*{-1.5cm}
\centerline{\large EUROPEAN ORGANIZATION FOR NUCLEAR RESEARCH (CERN)}
\vspace*{1.5cm}
\hspace*{-0.5cm}
\begin{tabular*}{\linewidth}{lc@{\extracolsep{\fill}}r}
\vspace*{-1.2cm}\mbox{\!\!\!\includegraphics[width=.12\textwidth]{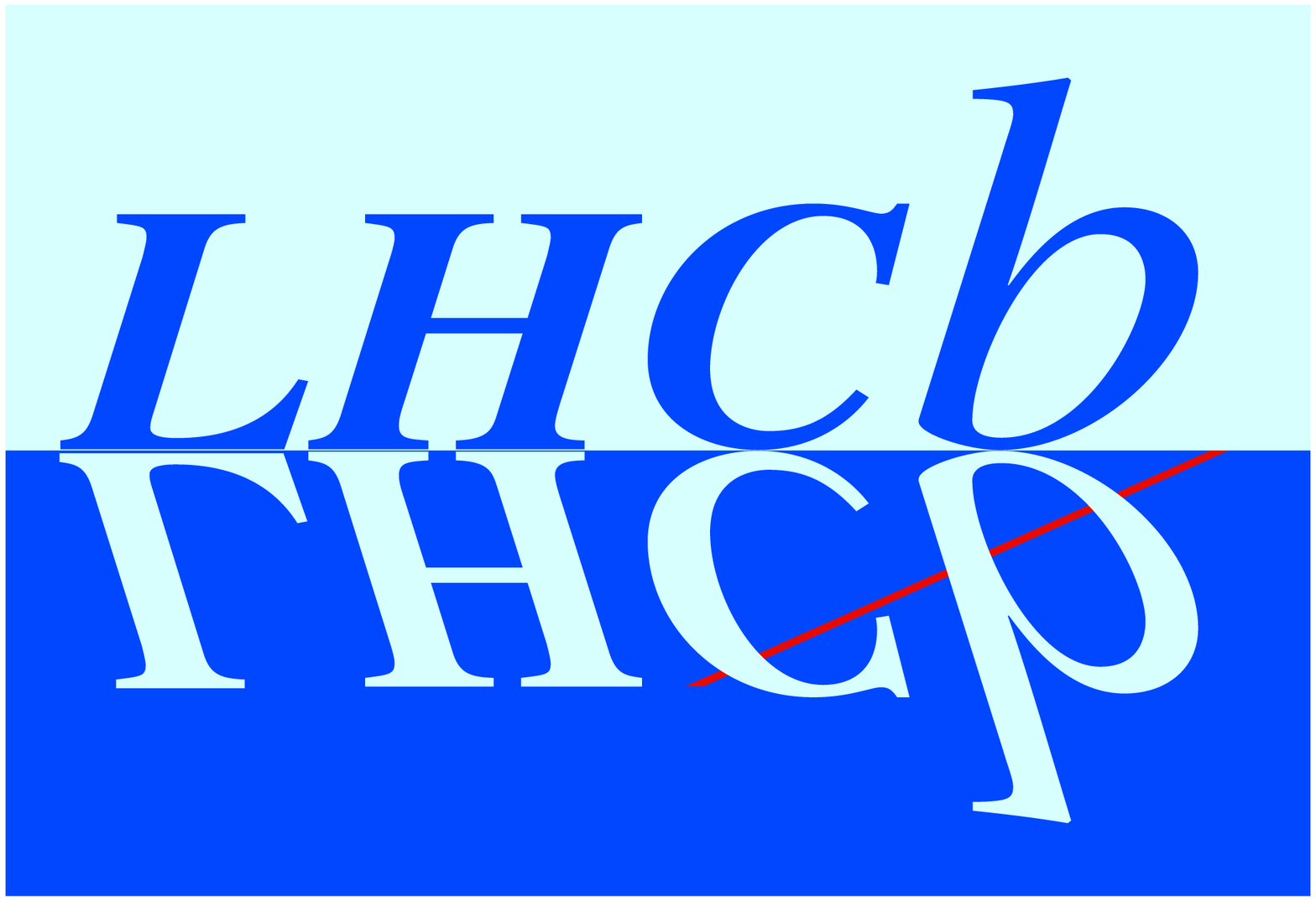}} & &%
\\
 & & CERN-PH-EP-2013-025 \\  
 & & LHCb-PAPER-2012-056 \\  
 & & February 26, 2013 \\
\end{tabular*}

\vspace*{2.0cm}

{\bf\boldmath\huge
\begin{center}
Search for the decay $B_s^0 \to D^{*\mp} \pi^\pm$
\end{center}
}
\vspace*{1.0cm}

\begin{center}
The LHCb collaboration\footnote{Authors are listed on the following pages.}
\end{center}\

\vspace{\fill}

\begin{abstract}
\noindent
A search for the decay $B_s^0 \to D^{*\mp} \pi^\pm$ is presented using a data sample corresponding to an integrated luminosity of $1.0 \ {\rm fb}^{-1}$ of $pp$ collisions collected by LHCb.
This decay is expected to be mediated by a $W$-exchange diagram, with little contribution from rescattering processes, and therefore a measurement of the branching fraction will help to understand the mechanism behind related decays such as $B_s^0 \to \pi^+\pi^-$ and $B_s^0 \to \D\Dbar$.
Systematic uncertainties are minimised by using $\Bd \to \Dstarmp\pipm$ as a normalisation channel. 
We find no evidence for a signal, and set an upper limit on the branching fraction of 
${\cal B}(B_s^0 \to D^{*\mp} \pi^\pm) < 6.1\,\left(7.8\right) \times 10^{-6}$ at 90\,\% (95\,\%) confidence level. 
\end{abstract}

\vspace*{1.0cm}
\begin{center}
Submitted to Phys.\ Rev.\ D (R)
\end{center}

\vspace{\fill}

{\footnotesize
 \centerline{\copyright~CERN on behalf of the \lhcb collaboration, license \href{http://creativecommons.org/licenses/by/3.0/}{CC-BY-3.0}.}}
\vspace*{2mm}

\end{titlepage}

\newpage
\setcounter{page}{2}
\mbox{~}
\newpage

\centerline{\large\bf LHCb collaboration}
\begin{flushleft}
\small
R.~Aaij$^{40}$, 
C.~Abellan~Beteta$^{35,n}$, 
B.~Adeva$^{36}$, 
M.~Adinolfi$^{45}$, 
C.~Adrover$^{6}$, 
A.~Affolder$^{51}$, 
Z.~Ajaltouni$^{5}$, 
J.~Albrecht$^{9}$, 
F.~Alessio$^{37}$, 
M.~Alexander$^{50}$
% ...

{\footnotesize \it
% ...
$ ^{5}$Clermont Universit\'{e}, Universit\'{e} Blaise Pascal, CNRS/IN2P3, LPC, Clermont-Ferrand, France\\
$ ^{6}$CPPM, Aix-Marseille Universit\'{e}, CNRS/IN2P3, Marseille, France\\
% ...
$ ^{9}$Fakult\"{a}t Physik, Technische Universit\"{a}t Dortmund, Dortmund, Germany\\
% ...
$ ^{35}$Universitat de Barcelona, Barcelona, Spain\\
$ ^{36}$Universidad de Santiago de Compostela, Santiago de Compostela, Spain\\
$ ^{37}$European Organization for Nuclear Research (CERN), Geneva, Switzerland\\
% ...
$ ^{40}$Nikhef National Institute for Subatomic Physics, Amsterdam, The Netherlands\\
% ...
$ ^{45}$H.H. Wills Physics Laboratory, University of Bristol, Bristol, United Kingdom\\
% ...
$ ^{50}$School of Physics and Astronomy, University of Glasgow, Glasgow, United Kingdom\\
$ ^{51}$Oliver Lodge Laboratory, University of Liverpool, Liverpool, United Kingdom\\
% ...
$ ^{n}$LIFAELS, La Salle, Universitat Ramon Llull, Barcelona, Spain\\
% ...
}
\end{flushleft}

\cleardoublepage

\renewcommand{\thefootnote}{\arabic{footnote}}
\setcounter{footnote}{0}


\pagestyle{plain} 
\setcounter{page}{1}
\pagenumbering{arabic}

\begin{figure}[!htb]
\centering
\includegraphics[width=0.31\textwidth]{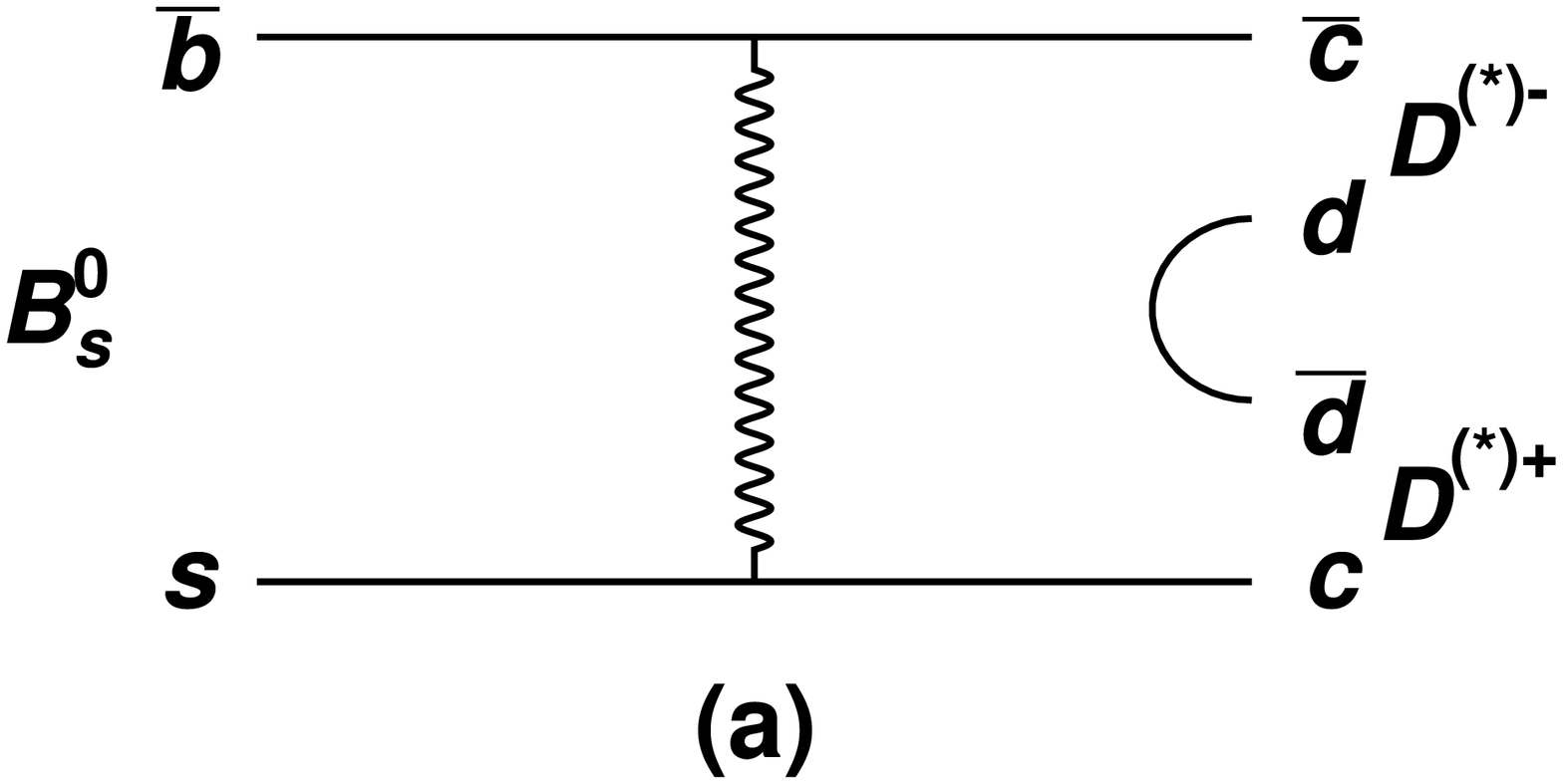}
\hspace{2mm}
\includegraphics[width=0.31\textwidth]{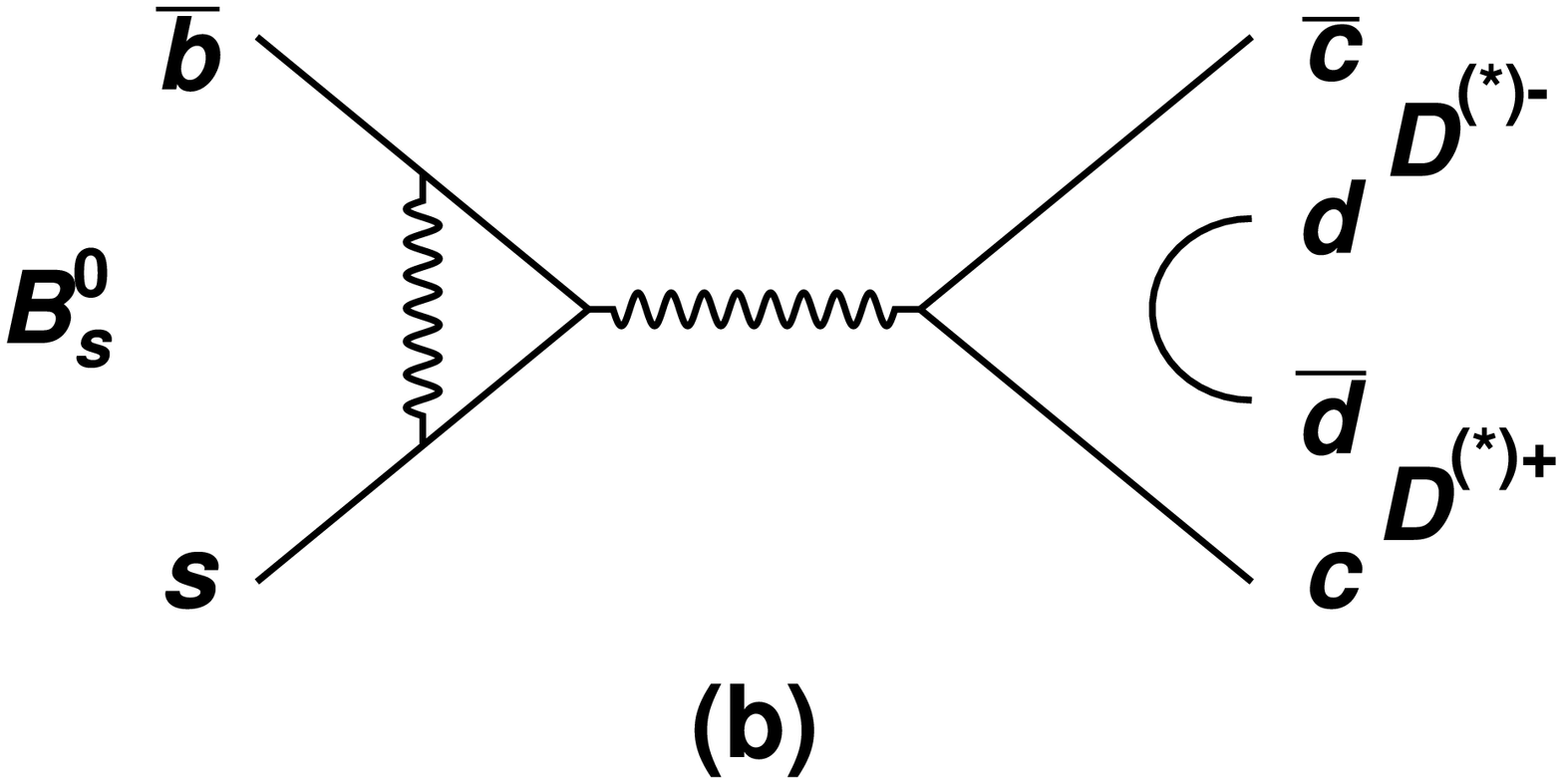}
\hspace{2mm}
\includegraphics[width=0.31\textwidth]{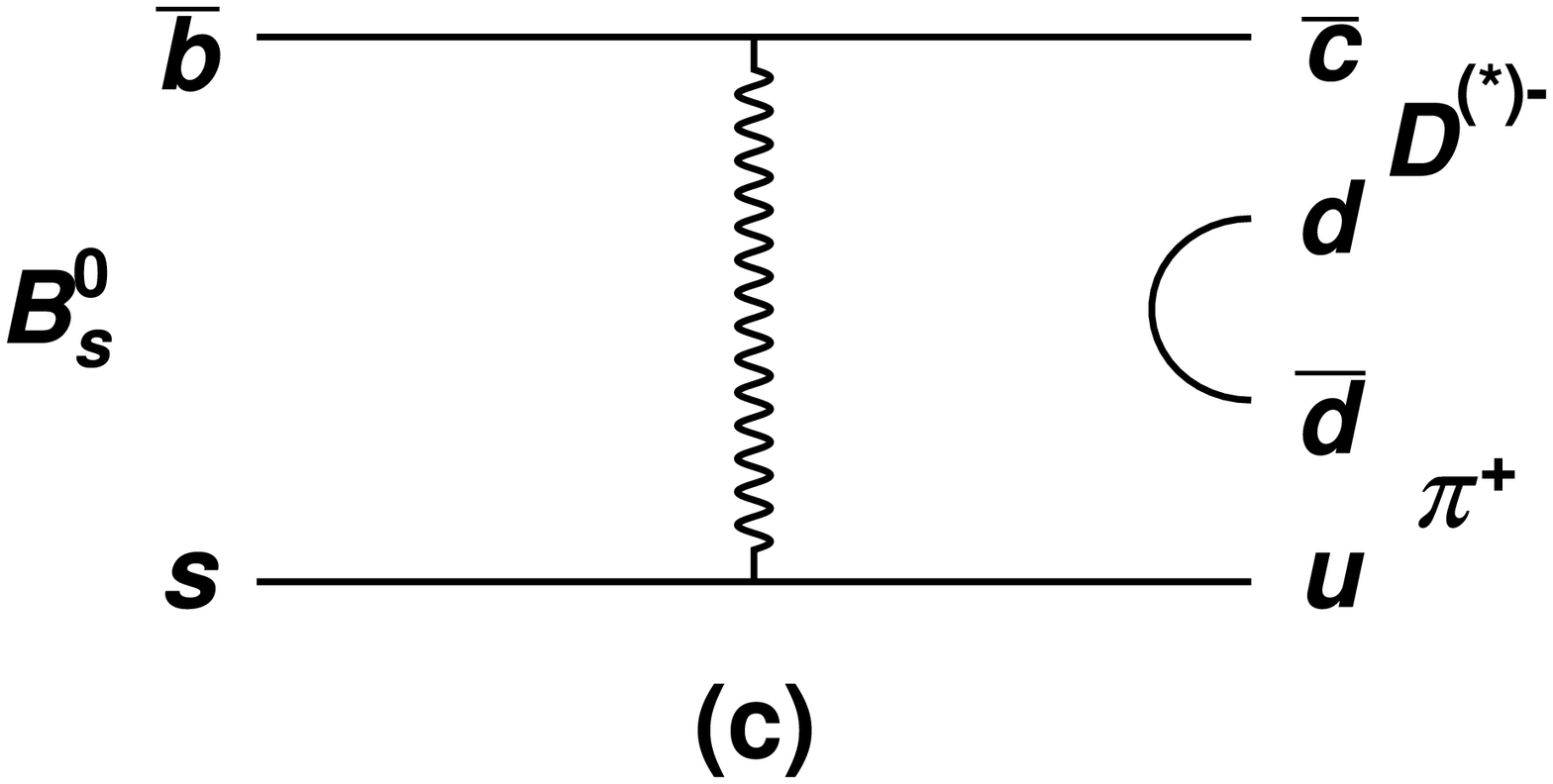}
\caption{\small
Decay diagrams for (a) $\Bs \to D^{(*)+}D^{(*)-}$ via weak exchange, (b) $\Bs \to D^{(*)+}D^{(*)-}$ via penguin annihilation, (c) $\Bs \to D^{(*)-}\pip$ via weak exchange.}
\label{fig:feynman}
\end{figure}

Decays of \Bs mesons to final states such as $\Dp\Dm$, $\Dz\Dzb$~\cite{LHCb-PAPER-2012-050} and $\pipi$~\cite{LHCb-PAPER-2012-002} have been recently observed by LHCb.
Such decays can proceed, at short distance, by two types of amplitudes, referred to as weak exchange and penguin annihilation.
Example diagrams are shown in Fig.~\ref{fig:feynman}(a) and (b).
There is also a potential long distance contribution from rescattering.
For example, the $\Dp\Dm$ final state can be obtained from a $b \to c\bar{c}s$ decay to $\Dsp\Dsm$ followed by the $s\bar{s}$ pair rearranging to $d\bar{d}$.
Understanding rescattering effects in hadronic $B$ meson decays is important in order to interpret various \CP-violating observables.

A measurement of the branching fraction of the decay $\Bs \to \Dstarm\pip$ can be used to disentangle the contributions from different decay diagrams and from rescattering~\cite{Gronau:1995hm,Fleischer:2003yb}.
This decay has only weak exchange contributions, as shown in Fig.~\ref{fig:feynman}(c).
(The suppressed diagram for $\Bs \to \Dstarp\pim$ is not shown.)
Moreover, rescattering contributions to the $\Bs \to D^{(*)\mp}\pipm$ decay are expected to be small~\cite{Gronau:2012gs}.
Therefore, if the observed branching fraction for the decay $\Bs \to \pip\pim$ is explained by rescattering, a low value of ${\cal B}(\Bs \to \Dstarm\pip) = (1.2 \pm 0.2) \times 10^{-6}$ is predicted~\cite{Gronau:2012gs}.
However, if short-distance amplitudes are the dominant effect in $\Bs \to \pip\pim$ and related decays, ${\cal B}(\Bs \to \Dstarm\pip)$ could be much larger.
The measured $\Bs \to \D\Dbar$~\cite{LHCb-PAPER-2012-050} and $\Bp \to \Ds\phi$~\cite{LHCb-PAPER-2012-025} rates are at the upper end of the expected range in the rescattering-based model, but further measurements are needed to establish whether long-distance processes are dominant in these hadronic $B$ decays.

In this paper, the result of a search for the decay $\Bs \to D^{*\mp}\pipm$ is presented. No previous measurements of this decay have been made.
The inclusion of charge conjugated processes is implied throughout the paper.
Since the flavour of the $\Bs$ meson at production is not tagged, the $\Dstarm\pip$ and $\Dstarp\pim$ final states are combined.
The analysis is based on a data sample corresponding to an integrated luminosity of $1.0 \invfb$ of LHC $pp$ collision data, at a centre-of-mass energy of $7 \tev$, collected with the LHCb detector during 2011.
In high energy $pp$ collisions all $b$ hadron species are produced, so the $\Bd \to \Dstarmp\pipm$ decay, with branching fraction ${\cal B}(\Bd \to \Dstarm\pip) = (2.76 \pm 0.13) \times 10^{-3}$~\cite{Aubert:2006cd,PDG2012}, is both a potentially serious background channel as well as the ideal normalisation mode for the measurement of the $\Bs$ branching fraction.

The \lhcb detector~\cite{Alves:2008zz} is a single-arm forward
spectrometer covering the \mbox{pseudorapidity} range $2<\eta <5$,
designed for the study of particles containing \bquark or \cquark
quarks. The detector includes a high precision tracking system
consisting of a silicon-strip vertex detector surrounding the $pp$
interaction region, a large-area silicon-strip detector located
upstream of a dipole magnet with a bending power of about
$4{\rm\,Tm}$, and three stations of silicon-strip detectors and straw
drift tubes placed downstream. The combined tracking system has
momentum resolution $\Delta p/p$ that varies from 0.4\,\% at 5\gevc to
0.6\,\% at 100\gevc, and impact parameter (IP) resolution of 20\mum for
tracks with high transverse momentum ($\pt$). Charged hadrons are identified
using two ring-imaging Cherenkov detectors. Photon, electron and
hadron candidates are identified by a calorimeter system consisting of
scintillating-pad and preshower detectors, an electromagnetic
calorimeter and a hadronic calorimeter. Muons are identified by a
system composed of alternating layers of iron and multiwire
proportional chambers.

The trigger~\cite{Aaij:2012me} consists of a hardware stage, based on information from the calorimeter and muon systems, followed by a software stage which applies a full event reconstruction.
In this analysis, signal candidates are accepted if one of the final state particles created a cluster in the calorimeter with sufficient transverse energy to fire the hardware trigger.
Events that are triggered at the hardware level by another particle in the $pp \to b\bar{b}X$ event are also retained.
The software trigger requires characteristic signatures of $b$-hadron decays: at least one track, with $\pt > 1.7\,\gevc$ and $\chisq_{\rm IP}$ with respect to any primary interaction vertex (PV) greater than 16, that subsequently forms a two-, three- or four-track secondary vertex with a high sum of the $\pt$ of the tracks and significant displacement from the PV. The $\chisq_{\rm IP}$ is the difference between the \chisq of the PV reconstruction with and without the considered track.
In the offline analysis, the software trigger decision is required to be due to the candidate signal decay.

Candidates that are consistent with the decay chain $\Bds \to \Dstarmp\pipm$, $\Dstarm\to\Dzb\pim$, $\Dzb\to\Kp\pim$ are selected.
The $\Dzb$ and $\Dstarm$ candidate invariant masses are required to satisfy $1814< m_{\Kp\pim} < 1914 \mevcc$ and $2008.78 < m_{\Dzb\pim} < 2011.78 \mevcc$, respectively, where a $\Dz$ mass constraint is applied in the evaluation of $m_{\Dzb\pim}$.
The bachelor pion, from the \Bds decay, is required to be consistent with the pion mass hypothesis, based on particle identification (PID) information from the RICH detectors~\cite{arXiv:1211-6759}.
All other selection criteria were tuned on the $\Bd\to\Dstarmp\pipm$ control channel in a similar manner to that used in another recent LHCb publication~\cite{LHCb-PAPER-2012-018}.
The large yield in the normalisation sample allows the selection to be based on data, though the efficiencies are determined using Monte Carlo (MC) simulated events in which $pp$ collisions are generated using \pythia~6.4~\cite{Sjostrand:2006za} with a specific \lhcb configuration~\cite{LHCb-PROC-2010-056}.  Decays of hadronic particles are described by \evtgen~\cite{Lange:2001uf}. The interaction of the generated particles with the detector and its response are implemented using the \geant toolkit~\cite{Allison:2006ve, *Agostinelli:2002hh} as described in Ref.~\cite{LHCb-PROC-2011-006}.

The selection requirements include criteria on the quality of the tracks forming the signal candidate, their $p$, $\pt$ and inconsistency with the hypothesis of originating from the PV ($\chisq_{\rm IP}$). Requirements are also placed on the corresponding variables for candidate composite particles ($\Dzb$, $\Bds$) together with restrictions of the decay fit ($\chisq_{\rm vertex}$), the flight distance ($\chisq_{\rm flight}$), and the cosine of the angle between the momentum vector and the line joining the PV to the $\Bds$ vertex ($\cos \theta_{\rm dir}$)~\cite{LHCB-PAPER-2011-008}.

Further discrimination between signal and background categories is achieved by calculating weights for the remaining $\Bd$ candidates~\cite{Pivk:2004ty}. The weights are based on a simplified fit to the $B$ candidate invariant mass distribution, where the $\Bs$ region is neither examined nor included in the fit.
The weights are used to train a neural network~\cite{Feindt:2006pm} in order to maximise the separation between categories.
To retain sufficient background events for the network training, the requirement on $m_{\Dzb\pim}$ is not applied.
A total of fifteen variables are used as input to the network.
They include the $\chisq_{\rm IP}$ of the four candidate tracks, the $\chisq_{\rm IP}$, $\chisq_{\rm vertex}$, $\chisq_{\rm flight}$ and $\cos \theta_{\rm dir}$ of the $\Dzb$ and $\Bds$ candidates, and the $\Bds$ candidate $\pt$.
The $\pt$ asymmetry and track multiplicity in a cone with half-angle of 1.5 units in the plane of pseudorapidity and azimuthal angle (measured in radians)~\cite{LHCB-PAPER-2012-001} around the $\Bds$ candidate flight direction are also used.
The input quantities to the neural network depend only weakly on the kinematics of the $\Bds$ decay.
A requirement on the network output is imposed that reduces the combinatorial background by an order of magnitude while retaining about 75\,\% of the signal.
Potential biases from this data-driven method are investigated by training the neural network with different fractions of the data sample. The same results are obtained using a neural network trained on 30, 40, 50, 60 and 70\,\% of the total data sample.

After all selection requirements are applied, approximately 50\,000 candidates are selected in the invariant mass range $5150 < m_{\Dstarm\pip} < 5600 \mevcc$.
About 1\,\% of events with at least one candidate also contain a second candidate.
Such multiple candidates are retained and treated the same as other candidates.

In addition to combinatorial background, candidates may be formed from misidentified or partially reconstructed $\Bds$ decays.
Contributions from partially reconstructed decays are reduced by requiring the invariant mass of the $\Bds$ candidate to be above $5150 \mevcc$.
The contribution from $\Bds$ decays to identical final states but without intermediate charmed mesons is negligible due to the requirement on the $\Dstarm$ candidate invariant mass.
A small but significant number of background events are expected from $\Bd \to \Dstarm\Kp$ decays with the $\Kp$ misidentified as a pion.
The branching fractions of $\Bsb \to \Dstarm\Kp$ and $\Lb \to \Dstarm\proton$ are expected to be small due to CKM suppression, so that these potential backgrounds are negligible.

Since the \Bd decay mode is several orders of magnitude more abundant than the \Bs decay, it is critical to understand precisely the shape of the \Bd signal peak.
The dependence of the width of the peak on different kinematic variables of the \Bd decay was investigated.
The strongest correlation was found to be with the angle between the momenta of the \Dstarm candidate and the bachelor \pip in the lab frame, denoted as $\theta_{\rm bach}$.
Simulated pseudo-experiments were used to find an optimal number of $\theta_{\rm bach}$ bins to be used in a simultaneous fit.
The outcome is that five bins are used, with ranges 0--0.046, 0.046--0.067, 0.067--0.092, 0.092--0.128 and 0.128--0.4 \rad, chosen to have approximately equal numbers of $\Bd$ decays in each.
The peak width in the highest bin is approximately 60\,\% of that in the lowest bin.
The pseudo-experiments show that the simultaneous fit in bins of $\theta_{\rm bach}$ is approximately 20\,\% more sensitive to a potential $\Bs$ signal than the fit without binning.

The signal yields are obtained from a maximum likelihood fit to the $\Dstarm\pip$ invariant mass distribution in the range $5150$--$5600 \mevcc$. The fit is performed simultaneously in the five $\theta_{\rm bach}$ bins.
The fit includes double Gaussian shapes, where the two Gaussian functions share a common mean, for \Bd and \Bs signals, together with an exponential component for the partially reconstructed background, a linear component for the combinatorial background and a non-parametric function, derived from simulation, for $\Bd\to\Dstarm\Kp$ decays.
The probability density function (PDF) for the $\Bd\to\Dstarm\Kp$ background is shifted by the mass difference between data and simulation for each bin of $\theta_{\rm bach}$.

The parameters of the double Gaussian shapes are constrained to be identical for $\Bd$ and $\Bs$ signals, with an offset in their mean values fixed to the known $\Bd$--$\Bs$ mass difference~\cite{PDG2012}.
Additionally, the relative normalisation of the two Gaussian functions and the ratio of their widths are constrained within uncertainties to the value obtained in simulation.
A total of thirty-three parameters are allowed to vary in the fit: the ratio of yields $N(\Bs)/N(\Bd)$,
the linear slope of the combinatorial background and the exponential parameter of the partially reconstructed background, plus separate parameters in each of the $\theta_{\rm bach}$ bins to describe the peak position and core Gaussian width of the signal PDF, and the yields of the $\Bd$ peak, the combinatorial background, the partially reconstructed background, and the background from $\Bd \to \Dstarm\Kp$.

\begin{figure}[!htb]
 \centering
 \includegraphics[scale=0.40]{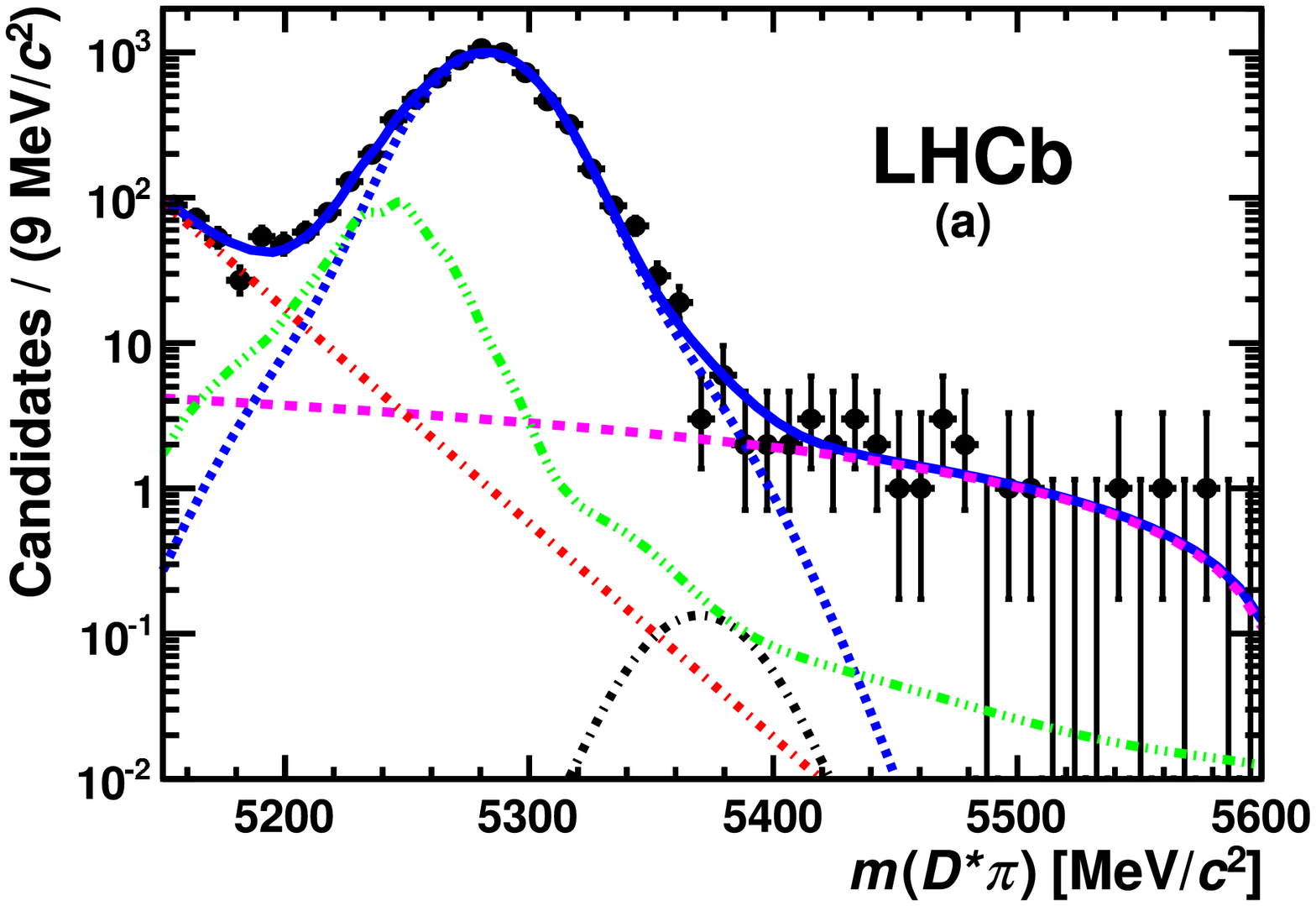}
 \includegraphics[scale=0.40]{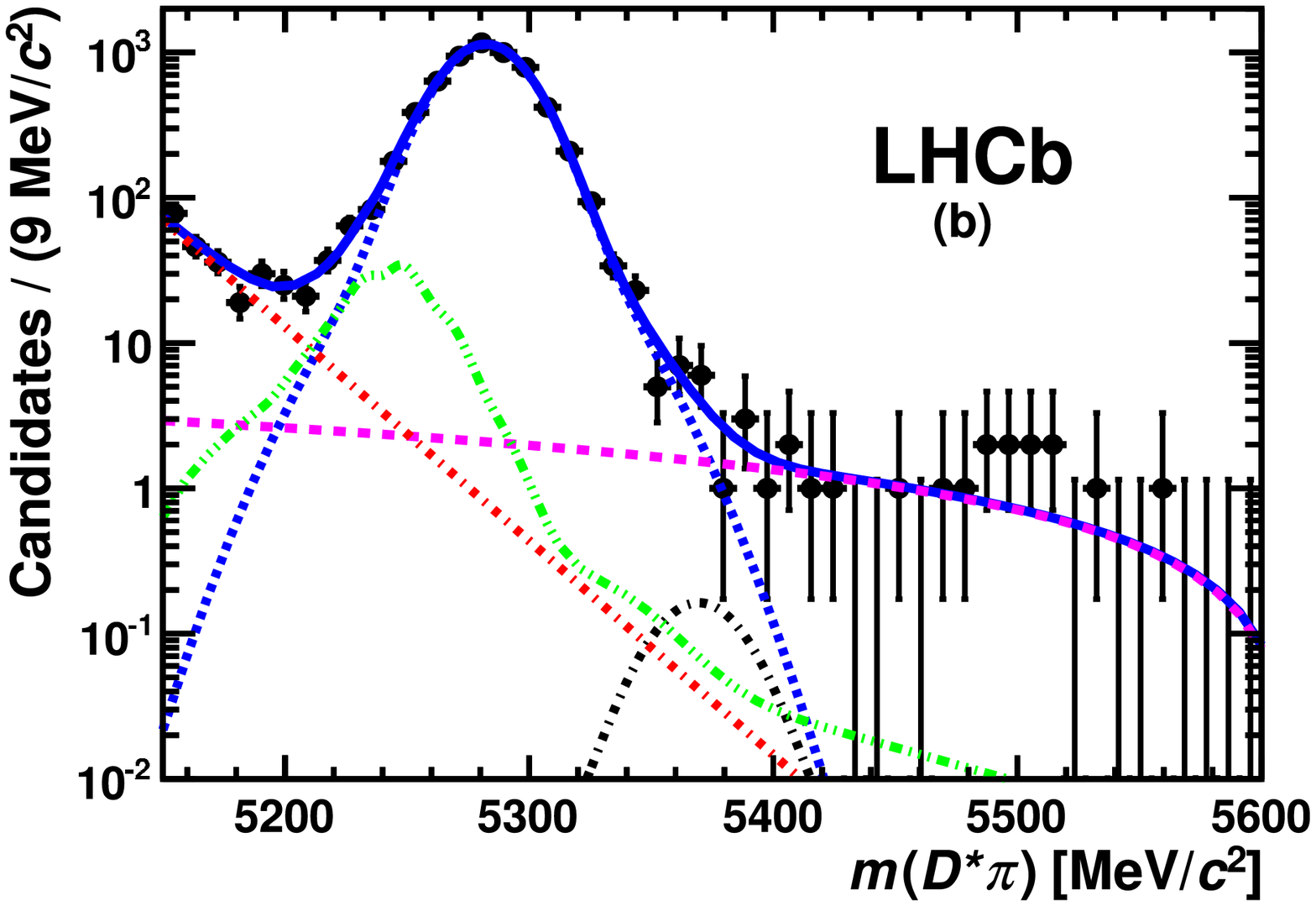}
 \vspace{2mm}
 \includegraphics[scale=0.40]{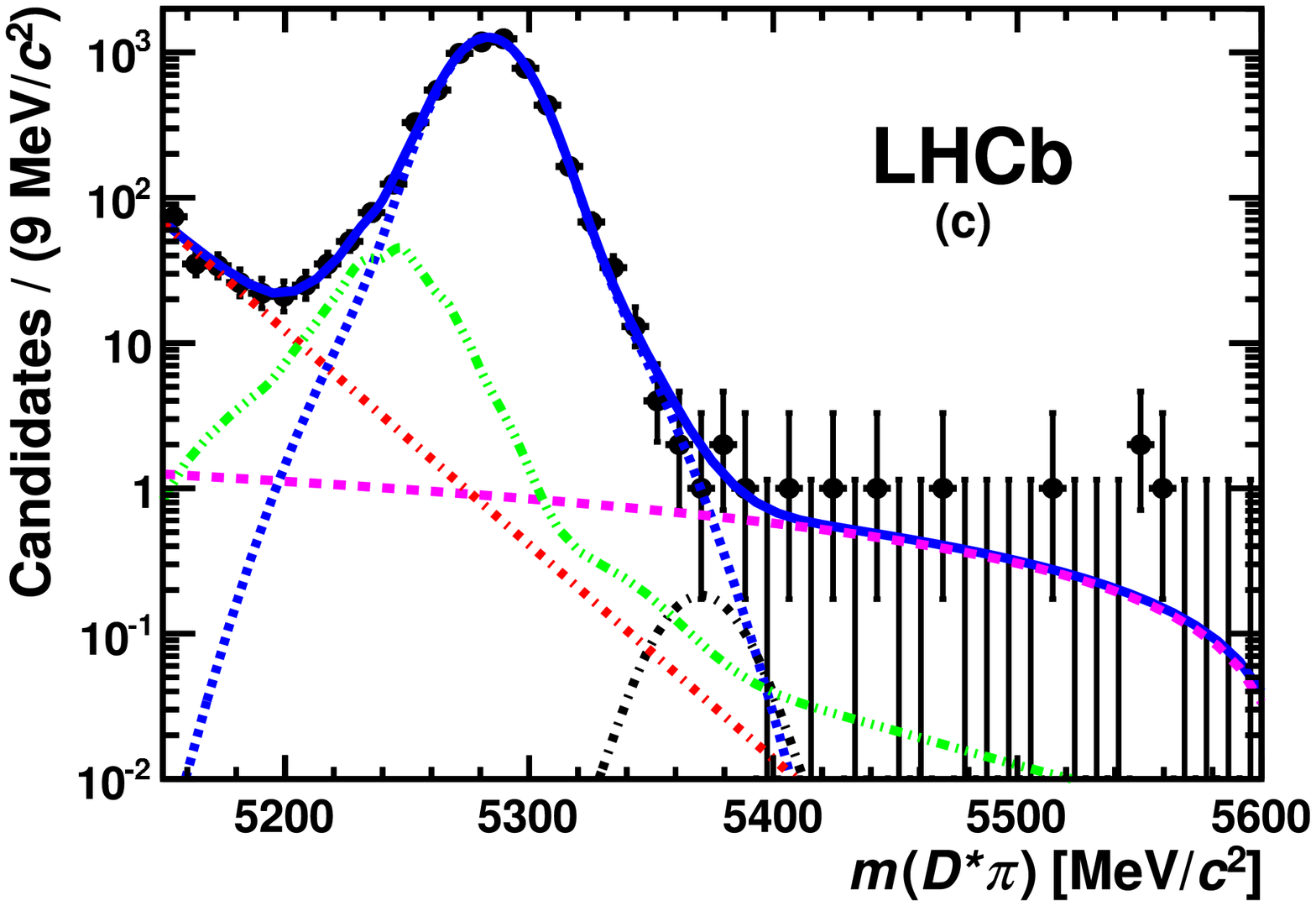}
 \includegraphics[scale=0.40]{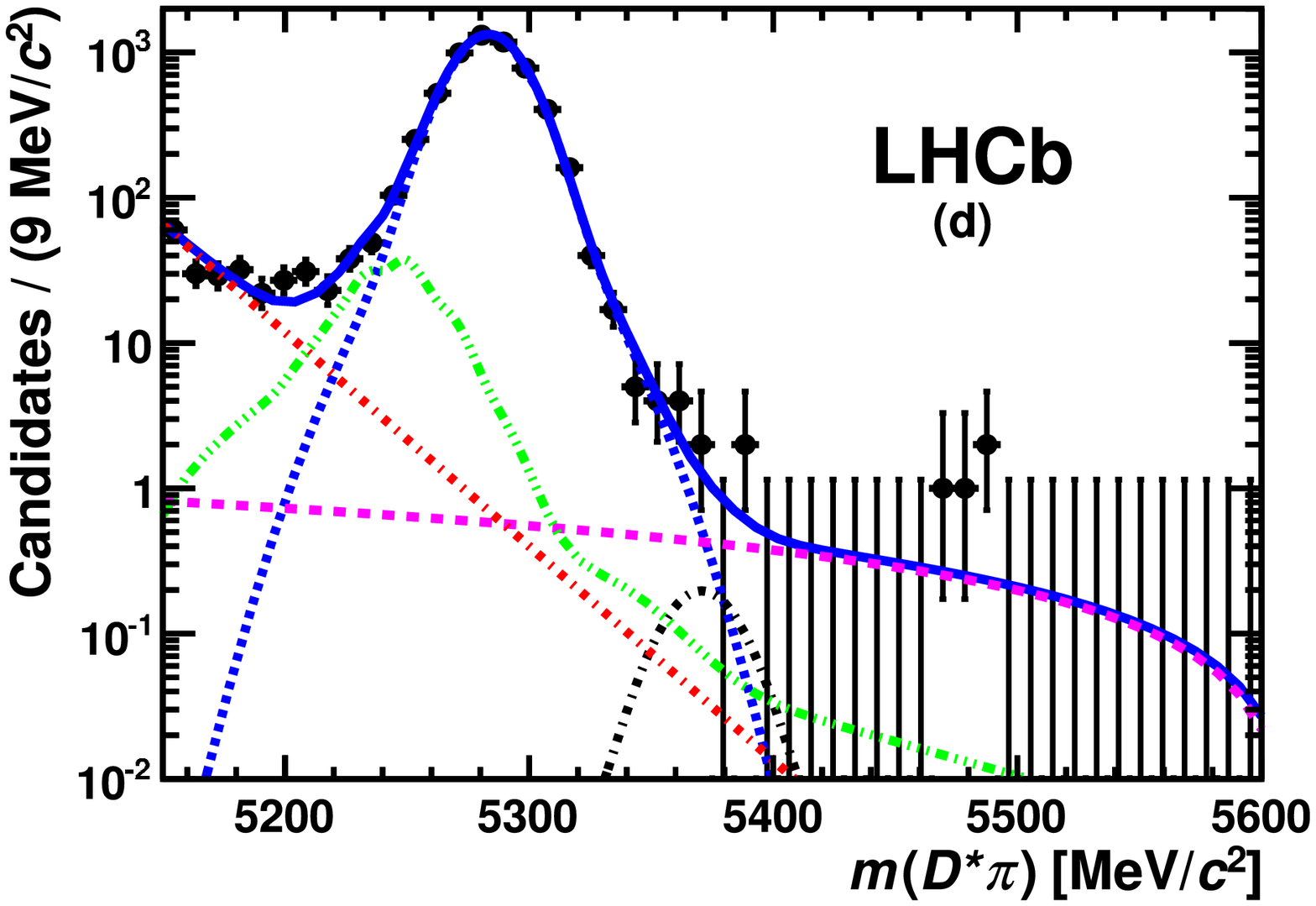}
 \vspace{2mm}
 \includegraphics[scale=0.40]{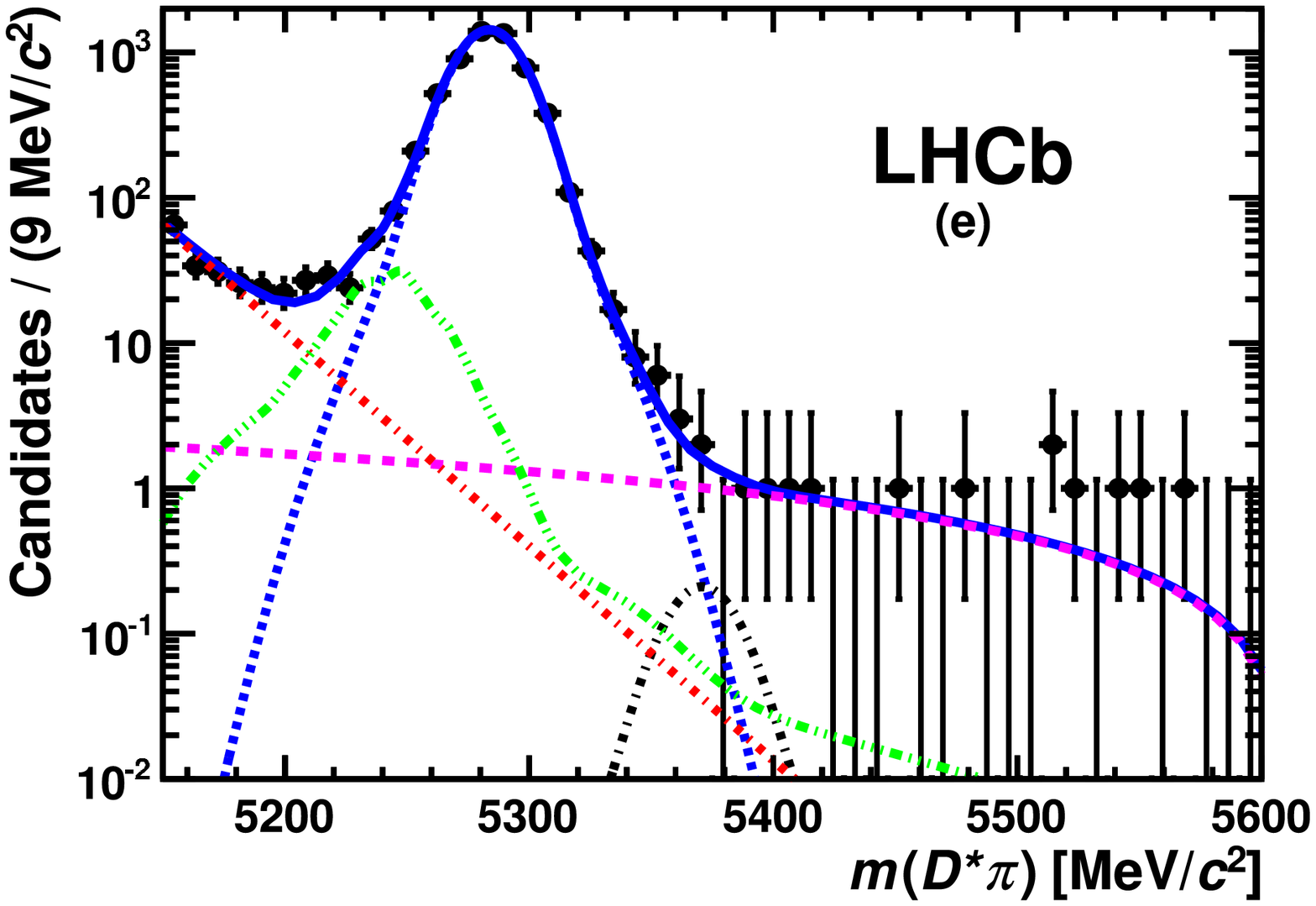}
 \includegraphics[scale=0.40]{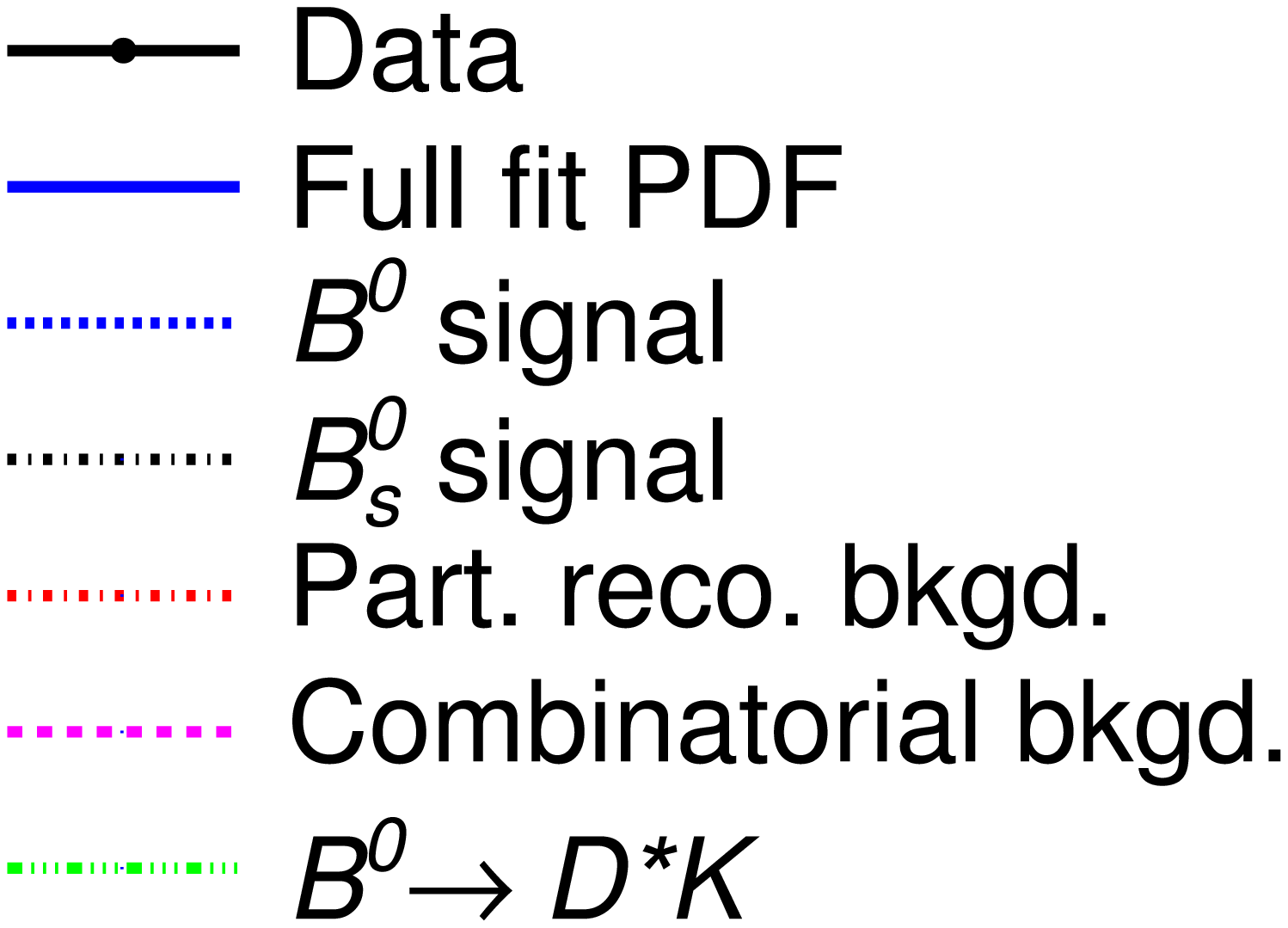}
\caption{\small
Simultaneous fit to the full data sample in five bins of $\theta_{\rm bach}$: (a) 0--0.046, (b) 0.046--0.067, (c) 0.067--0.092, (d) 0.092--0.128 and (e) 0.128--0.4 \rad. Note the $y$-axis scale is logarithmic and is the same for each bin. Data points are shown in black, the full PDF as a solid blue line and the component PDFs as: (red dot-dashed) partially reconstructed background, (magenta dashed) combinatorial background, (blue dashed) \Bd signal, (black dot-dashed) \Bs signal and (green 3 dot-dashed) $\Bd\to\Dstarm\Kp$ background.}
\label{fig:result2}
\end{figure}

The results of the fit are shown in Fig.~\ref{fig:result2}.
The total number of $\Bd \to \Dstarmp\pipm$ decays is found to be $30\,000 \pm 400$, and the ratio of yields is determined to be $N(\Bs)/N(\Bd) = (1.4 \pm 3.5) \times 10^{-4}$, where the uncertainty is statistical only.
The number of $\Bd \to \Dstarm\Kp$ decays found is $1\,200 \pm 200$, with a correlation of $7\,\%$ to the ratio of signal yields.

The ratio of yields is converted to a branching fraction following
\begin{equation}
{\cal B}(\Bs \to \Dstarmp\pipm) = \frac{N(\Bs)}{N(\Bd)} \times \frac{\epsilon(\Bd)}{\epsilon(\Bs)} \times \frac{f_d}{f_s} \times {\cal B}(\Bd \to \Dstarm\pip) \, ,
\end{equation}
where $\epsilon(\Bd)$ and $\epsilon(\Bs)$ are the efficiencies for the \Bd and \Bs decay modes respectively, while $f_d$ ($f_s$) is the probability that a \bquark quark produced in the acceptance results in a \Bd (\Bs) meson.
Their ratio has been determined to be $f_s/f_d = 0.256 \pm 0.020$~\cite{LHCb-PAPER-2012-037}.

The total efficiencies are $(0.165\pm0.002)\,\%$ and $(0.162\pm0.002)\,\%$ for the \Bd and \Bs decay modes respectively, including contributions from detector acceptance, selection criteria, PID and trigger effects.
The ratio is consistent with unity, as expected.
The PID efficiency is measured using a control sample of $\Dstarm \to \Dzb \pim,\,\Dzb \to \Kp\pim$ decays to obtain background-subtracted efficiency tables for kaons and pions as functions of their $p$ and $\pt$~\cite{LHCb-PAPER-2012-002}.
The kinematic properties of the tracks in signal decays are obtained from simulation, allowing the PID efficiency for each event to be obtained from the tables. Note that this calibration sample is dominated by promptly produced \Dstar mesons.
The remaining contributions to the total efficiency are determined from simulation, and validated using data.

Systematic uncertainties on ${\cal B}(\Bs \to \Dstarmp\pipm)$ are assigned due to the following sources, given in units of $1 \times 10^{-6}$, summarised in Table~\ref{tab:syst}.
\begin{table}
\centering
\caption{\small
Systematic uncertainties on ${\cal B}(\Bs \to \Dstarmp\pipm)$.}
\label{tab:syst}
\begin{tabular}{lc}
\hline
Source & Uncertainty $(10^{-6})$ \\
\hline
Efficiency & 0.02 \\
Fit model & 1.44 \\
Fit bias & 0.12 \\
Multiple candidates & 0.22 \\
$f_s/f_d$ & 0.12 \\
${\cal B}(\Bd \to \Dstarm\pip)$ & 0.08 \\
\hline
Total & 1.47 \\
\hline
\end{tabular}
\end{table}
Event selection efficiencies for both modes are found to be consistent in simulation to within $2\,\%$, yielding a  systematic uncertainty of $0.02$.
The fit model is varied by replacing the double Gaussian signal shapes with double Crystal Ball~\cite{Skwarnicki:1986xj} functions (with both upper and lower tails), changing the linear combinatorial background shape to quadratic and including a possible contribution from $\Bsb\to\Dstarm\Kp$. The non-parametric function for the $\Bd\to\Dstarm\Kp$ background was scaled in each bin to account for the change in the width of the \Bd signal. Combined in quadrature these sources contribute $1.44$ to the systematic uncertainty.
Possible biases in the determination of the fit parameters are investigated by simulated pseudo-experiments, leading to an uncertainty of $0.12$.
Events with multiple candidates are investigated by performing a fit having chosen one candidate at random. This fit is performed 100 times, with different seeds, and the spread of the results, $0.22$, is taken as the systematic uncertainty.
The uncertainty on the quantity $f_s/f_d$ contributes $0.12$, while that on ${\cal B}(\Bd \to \Dstarm\pip)$ gives $0.08$.
Combining all sources in quadrature, the total absolute systematic uncertainty is $1.47\times 10^{-6}$, and the \Bs branching fraction is determined to be
${\cal B}(\Bs \to \Dstarmp\pipm) = (1.5 \pm 3.8 \pm 1.5) \times 10^{-6}$,
where the first uncertainty is statistical and the second is systematic.

A number of cross-checks are performed to test the stability of the result. Candidates are divided based upon the hardware trigger decision into three groups; events in which a particle from the signal decay created a large enough cluster in the calorimeter to fire the trigger, events that were triggered independently of the signal decay and those events that were triggered by both the signal decay and the rest of the event.
The neural network and PID requirements are tightened and loosened.
The non-parametric PDF used to describe the background from D*K decays was smoothed to eliminate potential statistical fluctuations.
All cross-checks give consistent results.

Since no significant signal is observed, upper limits are set, at both 90\,\% and 95\,\% confidence level (CL), using a Bayesian approach.
The statistical likelihood curve from the fit is convolved with a Gaussian function of width given by the systematic uncertainty, and the upper limits are taken as the values containing 90\,\% (95\,\%) of the integral of the likelihood in the physical region.
The obtained limits are
\begin{equation*}
{\cal B}(\Bs \to \Dstarmp\pipm) < 6.1 \ (7.8) \times 10^{-6} \ {\rm at} \ 90\,\% \ (95\,\%) \ {\rm CL} \, .
\end{equation*}

In summary, the decay $\Bs \to \Dstarmp\pipm$ is searched for in a data sample of $1.0 \invfb$ of data collected with the LHCb detector during 2011.
No significant signal is observed and upper limits on the branching fraction are set.
The absence of a detectable signal indicates that rescattering effects may make significant contributions to other hadronic decays, such as $\Bs \to \pip\pim$ and $\Bs \to \D\Dbar$, as recently suggested~\cite{Gronau:2012gs}.

\section*{Acknowledgements}

\noindent We express our gratitude to our colleagues in the CERN
accelerator departments for the excellent performance of the LHC. We
thank the technical and administrative staff at the LHCb
institutes. We acknowledge support from CERN and from the national
agencies: CAPES, CNPq, FAPERJ and FINEP (Brazil); NSFC (China);
CNRS/IN2P3 and Region Auvergne (France); BMBF, DFG, HGF and MPG
(Germany); SFI (Ireland); INFN (Italy); FOM and NWO (The Netherlands);
SCSR (Poland); ANCS/IFA (Romania); MinES, Rosatom, RFBR and NRC
``Kurchatov Institute'' (Russia); MinECo, XuntaGal and GENCAT (Spain);
SNSF and SER (Switzerland); NAS Ukraine (Ukraine); STFC (United
Kingdom); NSF (USA). We also acknowledge the support received from the
ERC under FP7. The Tier1 computing centres are supported by IN2P3
(France), KIT and BMBF (Germany), INFN (Italy), NWO and SURF (The
Netherlands), PIC (Spain), GridPP (United Kingdom). We are thankful
for the computing resources put at our disposal by Yandex LLC
(Russia), as well as to the communities behind the multiple open
source software packages that we depend on.


\ifx\mcitethebibliography\mciteundefinedmacro
\PackageError{LHCb.bst}{mciteplus.sty has not been loaded}
{This bibstyle requires the use of the mciteplus package.}\fi
\providecommand{\href}[2]{#2}


\end{document}